\documentclass[conference]{IEEEtran}
\IEEEoverridecommandlockouts{}
\usepackage{cite}
\usepackage{amsmath,amssymb,amsfonts}
\usepackage{algorithmic}
\usepackage{graphicx}
\usepackage[full]{textcomp}
\usepackage{xcolor}

\usepackage{booktabs} % Add
\usepackage{adjustbox} % Add
\usepackage{pgfplots} % Add
\pgfplotsset{compat=1.17}
\usetikzlibrary{
  arrows.meta,
  calc, chains,
  quotes,
  positioning,
  shapes.geometric,
  shapes.symbols,
  shapes.misc,
  shadows,
}
\usepgfplotslibrary{groupplots}
\usepackage[frozencache=true,cachedir=minted-cache]{minted}
\setminted{
    fontsize=\footnotesize,
    frame=single,
    obeytabs=true,
    tabsize=2,
}

\makeatletter
\let\@float@c@listing\@caption
\makeatother

\usepackage{newtxmath} %add
\usepackage[final]{microtype} %add
\usepackage{ifpdf} 
\usepackage{ifxetex}
\ifxetex
  \usepackage[no-math]{fontspec}
\else
  \usepackage[T1]{fontenc}
  \usepackage{newtxmath,newtxtext}
\fi

\def\BibTeX{{\rm B\kern-.05em{\sc i\kern-.025em b}\kern-.08em
    T\kern-.1667em\lower.7ex\hbox{E}\kern-.125emX}}

\begin{document}

\title{\textit{REST API Fuzzing by Coverage Level Guided Blackbox Testing}\\

%\thanks{Sponsored in Part by the MOST, TWISC, and MOE.}
}

\author{
    \IEEEauthorblockN{
        Chung-Hsuan Tsai\IEEEauthorrefmark{1},
        Shi-Chun Tsai\IEEEauthorrefmark{1}, and
        Shih-Kun Huang\IEEEauthorrefmark{1}\IEEEauthorrefmark{2}
    }
    \IEEEauthorblockA{
        \textit{
            \IEEEauthorrefmark{1}Department of Computer Science,
            \IEEEauthorrefmark{2}Information Technology Service Center
        } \\
        \textit{
            National Yang Ming Chiao Tung University
        }\\
        Hsinchu, Taiwan \\
        \{zx.c, sctsai, skhuang\}@nycu.edu.tw
    }
}

\maketitle

\begin{abstract}
With the growth of web applications, REST APIs have become the primary communication method between services. In order to ensure system reliability and security, software quality can be assured by effective testing methods. Black box fuzz testing is one of the effective methods to perform tests on a large scale. However, conventional black box fuzz testing generates random data without judging the quality of the input.

We implement a black box fuzz testing method for REST APIs. It resolves the issues of blind mutations without knowing the effectiveness by Test Coverage Level feedback. We also enhance the mutation strategies by reducing the testing complexity for REST APIs, generating more appropriate test cases to cover possible paths.

We evaluate our method by testing two large open-source projects and 89 bugs are reported and confirmed. In addition, we find 351 bugs from 64 remote API services in APIs.guru.

The work is in https://github.com/iasthc/hsuan-fuzz.

\end{abstract}

\section{Introduction}\label{chapter:introduction}

Due to the limitation to exhaust possible input values of the program, there still exists potential errors in normal operations, and even leads to security concerns in the program. Software testing can effectively reduce risks, but manual testing and writing test cases are time-consuming, so we need to use automated tools to resolve this problem. Fuzz testing is one of the automated testing methods.

Fuzzing is very effective for triggering software errors. It can find exceptions that the developer has not dealt with, and identify whether the software has security issues. Fuzzing is also a common method of finding software vulnerabilities. For test targets with fixed binary formats (such as video and sound), fuzz testing is effective. There are many studies on black box, white box, or grey box test methods ~\cite{AFL++}~\cite{REDQUEEN}. However, network and Web applications are with variants of formats, and it is more difficult to use the same techniques as been used in fuzzing binary programs.

Internet applications and related services are popular. REST APIs have gradually become the primary communication method between services, ranging from cloud SaaS (software as a service) to IoT (Internet of Things) devices. If it can be tested on a large scale,  the quality of services will be improved. The emergence of the OpenAPI specification~\cite{openapi_specification} defines a standardized interface for REST APIs, making automated testing of a large number of REST APIs easier.

The current grey box fuzz testing in the REST API will track code coverage through program instrumentation, and guide the fuzzer to maximize code coverage, but it can only be applied to the programming language supported by the tool, such as Ruby~\cite{Pythia} or Python~\cite{shihtsunliu}. The black box fuzz testing will try to analyze and infer the dependencies between requests, and test by replacing fixed parameter values, as much as possible to combine the most API paths in each round of requests~\cite{RESTler}. Random tests may be carried out with the goal of increasing the coverage of status codes~\cite{EvoMaster2021}.

It can be seen from the above that grey box fuzzing can guide the fuzzing process through the feedback of coverage information, but it is limited by the specific programming languages supported. In contrast, although black box fuzzing has the advantage of not being restricted by the programming language, blind, random, and divergent mutations will make some paths hard to be triggered. In other words, if the black box fuzz testing process can identify the effects of the mutation, it can retain the advantages of not being restricted by the programming language and perform more efficient testing.

We, therefore, propose to add ``Test Coverage Level''~\cite{TCC} with feedback to the REST API black box fuzz testing to know whether the mutation effect and input are appropriate and to improve the shortcomings of the ordinary black box test that has no feedback from the testing results. Meanwhile, we resolve the issues that may be encountered in testing REST APIs and adopt different strategies to increase the speed and chance of triggering errors. Finally, a proof-of-concept tool is implemented that only requires OpenAPI specifications and path dependencies to perform automated testing, which can speed up the discovery of errors.

\begin{flushleft}
     \textbf{Our contributions:}
\end{flushleft}
\begin{itemize}
     \item Propose a new strategy for black box fuzzing with estimated code coverage.
     \item Implement a new black box fuzzer.
     \item Add ``Test Coverage Level'' as feedback for fuzzing.
     \item Resolve the issues of ``request sequence'', ``path dependency'', ``valid parameter'' and ``access token''.
     \item Use ``pairwise testing'' to reduce the combinations of test parameters and speed up the testing process.
     \item Increase the chance of triggering errors through different ``mutation strategies''.
\end{itemize}
\section{Background}\label{chapter:background}

We will explain ``OpenAPI Specification'', as well as ``Fuzz Testing'', ``Black Box Testing'', ``Pairwise Testing'' and ``Test Coverage Level'' respectively. 

\subsection{OpenAPI specification}

REST (Representational State Transfer) is a software architecture style for network applications. REST style is usually used in HTTP, mainly because the characteristics of HTTP are consistent with the definition of REST style, but it is not actually bound to any communication protocol. The Web API defined in this style is called RESTful API (hereinafter referred to as REST API).

\begin{table}[htbp]
    \caption{REST API and Non-REST API}\label{table:background_rest_non_rest}
    \begin{center}
        \begin{tabular}{cccc}
            \toprule
            \textbf{Behavior} & \textbf{Request Methods} & \multicolumn{1}{c}{\textbf{REST}} & \multicolumn{1}{c}{\textbf{Non-REST}} \\
            \midrule
            \midrule
            Create               & POST                    & /Users                            & /newUser                              \\
            \midrule
            Read              & GET                     & /Users/1                          & /getUser                              \\
            \midrule
            Update            & PUT                     & /Users/1                          & /updateUser                           \\
            \midrule
            Delete            & DELETE                  & /Users/1                          & /deleteUser                           \\
            \bottomrule
        \end{tabular}
    \end{center}
\end{table}

OpenAPI specification~\cite{openapi_specification} (hereinafter referred to as OpenAPI) is an interface description language, which is written in YAML or JSON format, not limited to a specific programming language so that REST API has a standardized description method. By directly reading the document or using the visualization tool~\cite{redoc}, you can learn the path, parameters, functions, and other information of the REST API, which is convenient for developers to implement and test according to the specifications.

\subsection{Fuzz testing}

Fuzz testing (Fuzzing) is a technique for automated software testing, which generates random and unexpected inputs repeatedly in the hope of triggering errors in the target program. It can effectively find program anomalies, logic errors, or developer design flaws, thereby improving program reliability and software quality. According to the obtained program information, testing can be divided into kinds of black box, white box, and grey box manner.

Black box testing usually refers to only understanding the program's input, output, and specifications, and not knowing the internal behavior of the program. Compared with black box testing, white box testing is with source code, which can generate better test cases through more complex techniques (such as symbolic execution), but it takes more time and cost.

Grey box testing is to have information about the specifications and program runtime information, such as tracking the code coverage achieved by each input through program instrumentation, understanding the effect of each round of mutation, and maximizing the code coverage. Most grey box fuzz testing~\cite{afl}~\cite{libFuzzer} adopt this scheme, and use better strategies such as seed selection and mutation methods to optimize fuzzing processes.

Code coverage is a software metric that indicates how much code is executed during the execution of a program. It provides an assessment and measure of the effectiveness of the tests. Compared to low code coverage, high code coverage has more chances for errors to occur.

\subsection{Pairwise testing}

Pairwise testing is combinatorial testing, dedicated to using fewer test cases to cover the paired combinations of multiple parameters. Since the program usually triggers the error~\cite{black2016pragmatic} by the interaction of single or paired parameters, this method can effectively reduce the number of tests. The minimum number of tests will be the product of the two most characteristic parameters.

\begin{center}
  \(C = \max_{i=1}^Q{B_i}\times\max_{j=1,j\neq i}^Q{B_j}\)
\end{center}

For example, if there are three different parameters of A, B, and C, with two, three, and four attributes respectively, the most comprehensive test combination needs to be executed 24 times (\(2\times 3 \times 4\)). If a paired test is used, Only need to execute 12 (\(3\times 4\)) times.

\subsection{Test Coverage Level}\label{section:background_TCL}

Martin-Lopez~\cite{TCC} et al.\ proposed a model for comparing test technologies for REST API.\@ By formulating ten test coverage criteria, and combining them into eight test coverage levels (hereinafter referred to as TCL) to overcome the inability to automatically measure the effectiveness of test cases. To achieve a specific TCL, all previous criteria must be met, which can be evaluated from different perspectives (such as API, path, or request method), and prove that TCL is correlated with code coverage and failure detection rate positively.

Table~\ref{table:background_test_coverage_criteria} is ``Test Coverage Model'' and the test coverage criteria that each TCL needs to include. We use TCL to be estimated code coverage and as feedback to the fuzzing process. 

\begin{table}[htbp]
	\caption{Test Coverage Model}\label{table:background_test_coverage_criteria}
	\begin{center}
		\begin{tabular}{cccc}
			\toprule
			\textbf{TCL} & \textbf{Input Criteria} & \textbf{Output Criteria} \\
			\midrule
			\midrule
			0            &                         &                          \\
			\midrule
			1            & Paths                   &                          \\
			\midrule
			2            & Operations              &                          \\
			\midrule
			3            & Content-type            & Content-type             \\
			\midrule
			4            & Parameters              & Status code classes      \\
			\midrule
			5            & Parameters              & Status codes             \\
			\midrule
			6            & Parameters              & Response body properties \\
			\midrule
			7            & Operation flows         &                          \\
			\bottomrule
		\end{tabular}
	\end{center}
\end{table}

\section{Design and Implementation}\label{chapter:implementation}

This research proposes a new concept ``REST API black box fuzz testing based on coverage level guidelines'', which uses TCL as feedback and guides the black box fuzzers to improve the drawback of black box testing without knowing the mutation effects. We summarize the problems encountered when testing REST APIs, and design our method to resolve the issues, and then explain our fuzzer ``HsuanFuzz''.

\subsection{Complexity of Testing}\label{section:complexity}

Functions in the application programs need to be called in the correct order, and REST API is no exception. Only by sending requests in the correct order can meaningful test cases be generated, thereby increasing code coverage. Test cases that are ``deleted'' and then ``updated'' are of little significance.

The highly structured REST API is also a problem. Paths are interdependent. The response of path ``A'' may be a parameter in the request of path ``B''. If the path dependency problem is not resolved, paths and methods cannot be tested completely, and ``404 Not Found'' responses will be received.

The format of parameters is also one of the issues to be resolved. Many parameter strings have specific formats, such as email, date or uuid, etc. Randomly generated values can easily lead to ``400 Bad Request''.

The last part is about authorization. API services obtain authorization methods in different ways, such as API key, Bearer token, or OAuth. If we do not obtain authorization correctly, we will get a response of ``401 Unauthorized'', and we will not be able to properly test the path that requires authentication.

Based on the above observations, four issues must be resolved: request sequence, path dependency, valid parameter, and access token.

\subsection{Architecture}

The purpose is to add a guideline method to the black box fuzzing to improve the disadvantage of not being able to determine the effect of the mutation. We input OpenAPI, path dependencies, and access information, and use TCL as feedback to automate the testing of REST API.\@ We also try to resolve the issues mentioned in Section~\ref{section:complexity}.

\begin{figure}[htbp]
	\centerline{
		\begin{tikzpicture}[
				node distance = 5mm and 6mm,
				base/.style = {draw,line width=0.2mm, minimum width=19.5mm, minimum height=6.5mm, align=center},
				startstop/.style = {base, rectangle, rounded corners},
				process/.style = {base, rectangle},
				io/.style = {draw,line width=0.2mm, minimum height=6.5mm, trapezium, trapezium left angle=70, trapezium right angle=110},
				decision/.style = {base, diamond},
				multidocument/.style={base, tape, draw, fill=white, tape bend top=none, double copy shadow},
				document/.style={base, tape, draw, tape bend top=none},
				input/.style={base,minimum width=8mm, minimum height=8mm, trapezium, draw, shape border rotate=90, trapezium left angle=90, trapezium right angle=80},
				datastore/.style={base, draw, minimum width=37mm, rounded rectangle, rounded rectangle east arc=concave, rounded rectangle arc length=150},
			every edge quotes/.style = {auto=right}]
			]
			
			\node (openapi) at (0, 0)     [io, align=center]                                {\footnotesize{OpenAPI, Path Dependencies,}\\ \footnotesize{and Access Information}};
			\node (grammar)               [process, below=of openapi]         {\footnotesize{Grammar}};
			\node (corpus)                [multidocument, below=of grammar]   {\footnotesize{Corpus}};
			\node (parser)                [document, below=of corpus]         {\footnotesize{Parser}};   
			\node (mutator)               [process, below=of parser]          {\footnotesize{Mutator}};
			\node (sender)                [process, below=of mutator]         {\footnotesize{Sender}};
			\node (analyzer)              [decision, below=of sender]         {\footnotesize{Analyzer}};
			\node (errors)                [multidocument, below=of analyzer]  {\footnotesize{Errors}};

			\draw [->, line width=0.2mm]  (corpus)    -- node[anchor=south,yshift=-0.2cm, xshift=0.8cm]{\footnotesize{Read}} (parser);
			\draw [->, line width=0.2mm]  (parser)    -- (mutator);
			\draw [->, line width=0.2mm]  (mutator)   -- (sender);
			\draw [->, line width=0.2mm]  (sender)    -- (analyzer);
			\draw [->, line width=0.2mm]  (analyzer)  -- node[anchor=south,yshift=0cm, xshift=2cm]{\footnotesize{500 Internal Server Error}} (errors);
			\draw [->, line width=0.2mm]  (grammar)   -- (corpus);
			\draw [->, line width=0.2mm]  (openapi)   -- (grammar);
			
			\draw [->, line width=0.2mm, dashed] (analyzer.west) 
			node[anchor=south,xshift=-1.2cm, yshift=-0.4cm,align=center]  {\footnotesize{TCL decrement}\\\footnotesize{or no changed}} -- ++(-2.5,0) |- 
			node[anchor=south,xshift=1.2cm, yshift=0cm]                   {\footnotesize{Next round}}
			(corpus.west);
			\draw [->, line width=0.2mm] (analyzer.east) 
			node[anchor=south, xshift=1.2cm,  yshift=0cm,align=center]  {\footnotesize{TCL increment}} -- ++(2.5,0) |-  
			node[anchor=south, xshift=-1.15cm, yshift=-0.05cm]                              {\footnotesize{Save to corpus}} 
			(corpus.east);
			
		\end{tikzpicture}
	}
	\caption{The API Fuzzing Architecture}\label{figure:implementation_flow}
\end{figure}
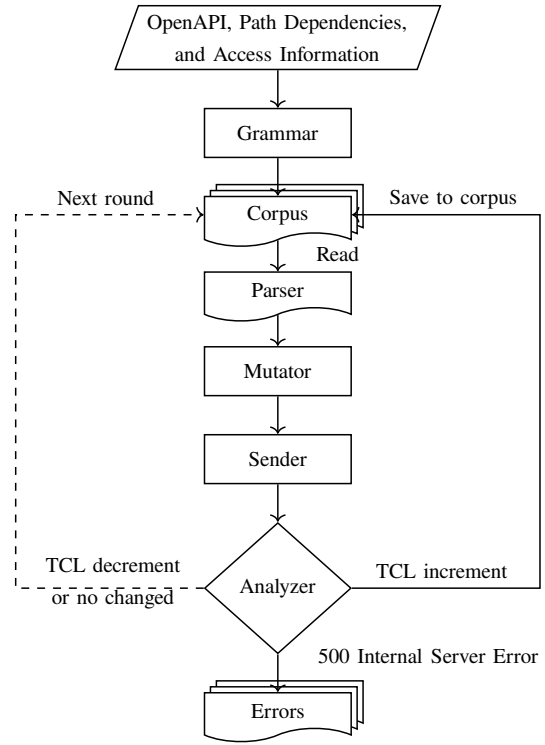

``Grammar'' phase in Fig.~\ref{figure:implementation_flow} analyzes the dependencies between OpenAPI and paths to ensure the order between paths and requests. To avoid incorrect parameter format, we use OpenAPI example values or pre-defined default values as initial values and store them in the corpus in the form of ProtoBuf.

After reading the seed, it needs to go through the ``parser'' phase to restore it to a grammar, replace the ID parameters to resolve the path dependence problem. The ``mutation'' phase mutates each request parameter in pairs. If authentication is required, we first obtain an access token and then send it along with the request in the ``sender'' phase. In addition, the response to a successful request is recorded for subsequent use.

In the ``analysis'' phase, the response of each path will be checked. If the TCL of any path increases, it will be added to the corpus. If the status code of any response is 500, an error will be recorded. The seeds are then read from the corpus and the steps above are repeated.

\subsection{Implementation}

We have implemented a proof-of-concept tool and named it ``HsuanFuzz'', which can perform black box fuzz testing based on coverage level guidelines on REST APIs. It can be executed by importing YAML files such as OpenAPI specifications, path dependencies, and access information (if required). We explain the details in the following.

\subsubsection{Setup of Initialization}

In order to make the fuzzer work correctly and efficiently, we need to enter some relevant information manually.

\paragraph{Entering related information}
``Path dependencies'' requires the tester to list the parameter dependencies of all paths, fill in the ID parameter and source of the request, and predefine the TCL operation flows, as shown in Listing~\ref{listing:implementation_dependencies_yaml}. 

In addition, the tester also needs to fill in the ``access information'' to obtain the token for REST API, or directly write the authorization key into the YAML file, as shown in Listing~\ref{listing:implementation_token_yaml}.

\paragraph{Defining Corpus and Error}

We use a map to create storage areas to avoid repeated storage of the same grammar or errors. The key is a hash of the path and request parameters, and the value is the serialized grammar. This allows us to restore the seed to grammar by deserialization.

\begin{listing}[H]
\begin{minted}{yaml}
paths:
  /articles/{slug}:
    items:
    - key: "slug"
      source:
        path: "/articles"
        key: "article[{slug}]"
posts:
  /articles:
    flow:
    - method: GET
      path: "/articles/{slug}"
    - method: PUT
      path: "/articles/{slug}"
\end{minted}
    \caption{Path Dependencies}
    \label{listing:implementation_dependencies_yaml}
\end{listing}

\begin{listing}[H]
\begin{minted}{yaml}
url: 'https://localhost/token'
method: POST
content_type: application/json
body: '{"username": "user", "password": "12345"}'
name: access_token
in: header
key: 'bearer'
token: ''
hardcode: false
\end{minted}
    \caption{Token Authorization}
    \label{listing:implementation_token_yaml}
\end{listing}

\subsubsection{Producing Seed Input} 

This phase has two parts. The first part is to generate the grammar, analyze the OpenAPI and solve the ``request sequence'' and ``path dependency'' problems. The second part is to assign parameter values, using OpenAPI examples or pre-defined default values.

\paragraph{Generating Grammar} 

The first execution will generate a grammar based on the defined data structure, and serialize it as a seed for the subsequent generation of more appropriate test cases. In order to analyze and calculate the TCL, we minimize the test cases, and each request will send the ``number of response status code types'' times.

\begin{listing}[H]
\begin{minted}{proto}
syntax = "proto3";

package base;
import "google/protobuf/struct.proto";

message Info {
    repeated Node nodes = 1;
}

message Node {
    uint32 group = 1;
    string path = 2;
    string method = 3;
    repeated Request requests = 4;
}

message Request {
    string type = 1;  
    google.protobuf.Struct value = 2;
}
\end{minted}
    \caption{Seed in ProtoBuf format}\label{figure:implementation_protobuf_proto}
\end{listing}

\paragraph{Solving dependencies}
The path itself has dependencies. It cannot be correctly deleted without creating an object. Paths are also dependent on each other. The value in the request parameter of path ``B'' needs to be obtained through the response of path ``A''. We have tried automated analysis, but found that the efficiency is low, because developers have different methods for defining parameters, and the user's ID parameter may be defined as a different name or even a typo, such as user\_id, userId, or usersId. Manual input combined with depth-first search can effectively improve the accuracy of dependencies. For example, if one wishes to test the path ``/articles/\{slug\}/comments'', the order of the requests will be the same as in Table~\ref{table:implementation_dependencies}, which can resolve the ``request order'' issue and overcome the ``path dependency'' problem.

\begin{table}[htbp]
	\caption{Request Order of Dependent Path}\label{table:implementation_dependencies}
	\begin{center}
		\begin{tabular}{ ccll }
			\toprule
			\textbf{*} & \multicolumn{1}{c}{\textbf{Method}} & \multicolumn{1}{c}{\textbf{Path}}  & \multicolumn{1}{c}{\textbf{Description}} \\
			\midrule
			\midrule
			1          & POST                                & /articles                          & New Article                              \\
			\midrule
			2          & PUT                                 & /articles/\{slug\}                 & Update an Article                        \\
			\midrule
			3          & GET                                 & /articles/\{slug\}                 & Read an Article                          \\
			\midrule
			4          & POST                                & /articles/\{slug\}/comments        & New Comment                              \\
			\midrule
			5          & GET                                 & /articles/\{slug\}/comments        & Read All Comments                        \\
			\midrule
			6          & DELETE                              & /articles/\{slug\}/comments/\{id\} & Delete a Comment                         \\
			\midrule
			7          & DELETE                              & /articles/\{slug\}                 & Delete an Article                        \\
			\bottomrule
		\end{tabular}
	\end{center}
\end{table}

\subsubsection{Giving Values to Parameters}

OpenAPI strings have many different formats. Randomly generating strings that do not conform to the format will result in ``400 Bad Request''. A good OpenAPI usually adds ``examples'' to the parameter attributes, so we can generate a valid initial value based on this information. If there is no example in OpenAPI, the initial value is generated according to the defined rules, such as Table~\ref{table:implementation_example_format}.

\begin{table}[htbp]
	\caption{Default Values of Parameter Format}\label{table:implementation_example_format}
	\begin{center}
		\begin{tabular}{ cl }
			\toprule
			\textbf{Format} & \multicolumn{1}{c}{\textbf{Default Values}} \\
			\midrule
			\midrule
			date            & 2021--05--28                                \\
			\midrule
			date-time       & 2021--05--28T10:00:00+08:00                 \\
			\midrule
			time            & 10:00:00+08:00                              \\
			\midrule
			email           & user@example.com                            \\
			\midrule
			hostname        & localhost                                   \\
			\midrule
			ipv4            & 127.0.0.1                                   \\
			\midrule
			ipv6            & 0:0:0:0:0:0:0:1                             \\
			\midrule
			uri             & https://tools.ietf.org/html/rfc3986         \\
			\midrule
			uuid            & 5bcafcb2--a669--11eb-bcbc--0242ac130002     \\
			\bottomrule
		\end{tabular}
	\end{center}
\end{table}

\subsection{Input Mutation}

The mutation phase will give priority to the path with the ID parameter, and then mutate according to the strategy. In order to avoid that the ID field can never be mutated, we put the mutation strategy behind it, that is, the ID parameters also have a certain chance to mutate to increase the chance of triggering errors.

\subsubsection{Handling ID Parameter} 

Correctly handling the ID parameters can greatly increase the number of successfully explored paths. We take the value from the response body through a defined method. Take Listing~\ref{listing:implementation_dependencies_yaml} as an example. ``key'' represents the ID parameter, and the ``path'' and ``key'' in the ``source'' are the data source of the ID parameter. For example, ``data[relationships\{id\}]'' represents the ``id'' field of the ``relationships'' object in the ``data'' array.

\subsubsection{Processing Mutation Strategy}

We use some methods of Go-fuzz~\cite{gofuzz} to modify the string, such as ``insert, duplicate, remove a range of bytes'', ``set, add, subtract a byte'', ``exchange two bytes'', ``replace a multi-byte ASCII number'' and ``bit flip''.
In addition, to modify the string directly, we added our own experience of writing REST API to the mutation strategy. If the exception is not handled properly, giving the wrong type or removing the necessary fields can easily cause the service to trigger an error. Therefore, according to the characteristics of the REST API, ``replace data type'', ``remove fields'' and ``modify string'' are randomly adopted.

However, mutating all parameters at the same time will easily lead to ``400 Bad Request''. Combining different numbers of parameters will cause the path to explode. Therefore, we adopt the method of pairwise testing. In each round of fuzzing, only two parameters of each request are changed to increase the chance and speed of triggering errors.

\subsection{Transferring Request}

After completing the pre-work, information on paths, methods, and parameters can be obtained from the mutated grammar. We set them to the correct fields, decode the string, remove control characters, and send the request after obtaining the access token. Lastly, we also record the response body of a successful request for using in the following requests.

\subsection{Analyzing Response}

In order to evaluate the effectiveness of the test case, we use the TCL mentioned in Section~\ref{section:background_TCL} as a feedback method and use the input and output criteria to calculate the TCL of each path. If the TCL of any path increases, it is defined as ``interesting'' and stored in the corpus for use in the next round of fuzzing. In addition, if the first digit of the response status code is five, it means that an error has been triggered, and will be recorded for later reproduction.

The maximum TCL of our implementation is 6 in general, unless there is a pre-defined operation flows ``Read Single'', ``Read All'', ``Update'', ``Delete'' in the input path dependencies  (Listing\ref{listing:implementation_dependencies_yaml}) and meet the behavior of Table~\ref{table:implementation_dependencies} to reach TCL 7.
\section{Results and Evaluations}\label{chapter:evaluation} 

We have implemented the fuzzer called HsuanFuzz. In order to be able to evaluate HsuanFuzz, based on the two different aspects of ``code coverage'' and ``error finding ability'', we discuss whether the research objectives are achieved in the form of research questions, and evaluate and present the results of this research.

\begin{itemize}
	\item RQ 1: Is it effective to add TCL (test coverage level) guidelines to the REST API black box fuzz testing?
	\item RQ 2: Is it better than existing REST API black box fuzzers?
	\item RQ 3: What other advantages does black box fuzzing have?
\end{itemize}

\subsection*{Experimental Environment}

Our devices are based on Ubuntu version 20.04 with Intel Core i7--10700 processor and 16 GB memory. We use ``HsuanFuzz based on black box coverage level guidelines'', ``HsuanFuzz without black box coverage level guidelines'' and black box fuzzer ``RESTler version 7.3.0''~\cite{restler_github} as comparisons. It is evaluated whether it achieves better results in different aspects according to the given time budget.

\subsection*{RQ 1: Guiding by TCL (Test Coverage Level) } 

In order to verify RQ 1, we selected the ``RealWorld'' program of Go language implementation~\cite{realworld_go} as the test target. We execute each fuzzer for 200 seconds and measure the code coverage with the number of lines of code executed as a unit.

\begin{table}[htbp]
	\caption{RealWorld Programs with Lines and Coverage}\label{table:evaluation_rq1_realworld_compare}
	\begin{center}
		\begin{tabular}{ccccc}
			\toprule
			\textbf{Components} & \textbf{Lines of Code} & \textbf{H\(^{\mathrm{a}}\)} & \textbf{H\(^{\mathrm{b}}\)} & \textbf{R\(^{\mathrm{c}}\)} \\
			\midrule
			\midrule
			User               & 319                    & 172                         & 163                         & 116                         \\
			\midrule
			Article          & 544                    & 438                         & 413                         & 36                          \\
			\midrule
			Total               & 863                    & 610                         & 576                         & 152                         \\
			\bottomrule
		\end{tabular}
	\end{center}
\end{table}

RealWorld~\cite{realworld} is a demonstration application that provides a complete API specification and can use different front-end and back-end implementations. There are two components: ``User'' and ``Article'', which can perform functions such as register, login, post articles, and comment. There are 11 paths and a total of 19 request methods.

{\let\thefootnote\relax\footnote{{
	\(^{\mathrm{a}}\) HsuanFuzz based on black box coverage level guidelines
}}}
{\let\thefootnote\relax\footnote{{
	\(^{\mathrm{b}}\) HsuanFuzz without black box coverage level guidelines
}}}
{\let\thefootnote\relax\footnote{{
	\(^{\mathrm{c}}\) RESTler
}}}

We use OpenAPI provided by RealWorld, and test in the default method of each fuzzer. Special attention is paid to the fact that OpenAPI does not provide parameter examples. HsuanFuzz can execute about 26 rounds during 200 seconds, which means that there are 26 chances to generate new seeds, while RESTler cannot define the number of requests for one round.

\begin{figure}[htbp]
	\centerline{
		\begin{tikzpicture}[scale=0.8]
			\begin{axis}[
					xlabel={Time (second)}, ylabel={Lines of Code},
					xmin=-10, xmax=210,
					legend style={
						at={(1.05, 1)},
						anchor=north west,
					},
					ymajorgrids=true,
					grid style=loosely dashed,
					line width=0.3mm
				]
				\addplot[
					color=blue,
					mark=none
				]
				coordinates {
					(0, 37) (1, 37) (1, 160) (2, 226) (3, 299) (3, 338) (4, 359) (5, 359) (6, 392) (6, 488) (7, 498) (8, 540) (8, 546) (9, 557) (10, 557) (11, 557) (12, 557) (12, 557) (13, 561) (14, 561) (15, 561) (15, 561) (16, 561) (57, 561) (58, 561) (59, 561) (60, 568) (61, 568) (62, 568) (63, 568) (63, 568) (64, 568) (65, 568) (66, 573) (67, 581) (68, 581) (69, 581) (70, 581) (71, 585) (72, 585) (72, 585) (74, 585) (75, 585) (76, 585) (77, 590) (78, 590) (78, 590) (79, 590) (80, 594) (81, 594) (82, 594) (84, 594) (85, 594) (86, 594) (87, 594) (87, 594) (89, 594) (90, 601) (91, 601) (93, 601) (94, 601) (95, 601) (95, 601) (97, 601) (98, 601) (100, 601) (100, 601) (102, 601) (102, 601) (104, 601) (105, 601) (107, 601) (108, 601) (109, 601) (110, 601) (111, 601) (112, 601) (114, 601) (116, 601) (157, 601) (158, 601) (160, 601) (162, 601) (163, 601) (165, 601) (166, 601) (167, 601) (169, 610) (171, 610) (172, 610) (173, 610) (175, 610) (177, 610) (178, 610) (180, 610) (181, 610) (182, 610) (184, 610) (186, 610) (187, 610) (188, 610) (189, 610) (192, 610) (194, 610) (195, 610) (196, 610) (198, 610) (200, 610)
				};
				
				\addplot[
					color=red,
					mark=none
				]
				coordinates {
					(0, 37) (0, 37) (1, 202) (2, 238) (2, 298) (3, 365) (4, 365) (5, 379) (5, 416) (6, 494) (7, 504) (7, 524) (48, 536) (49, 543) (49, 543) (50, 543) (51, 543) (52, 543) (53, 547) (53, 547) (54, 547) (55, 547) (56, 547) (56, 547) (57, 547) (58, 547) (59, 547) (60, 547) (60, 547) (61, 547) (62, 547) (63, 547) (64, 547) (65, 547) (66, 547) (66, 547) (67, 547) (68, 547) (69, 547) (70, 547) (71, 547) (72, 547) (72, 547) (73, 547) (75, 554) (76, 554) (77, 554) (77, 554) (78, 554) (79, 554) (80, 554) (81, 554) (82, 554) (83, 554) (84, 554) (85, 554) (86, 554) (87, 554) (88, 554) (89, 554) (90, 554) (91, 554) (92, 554) (94, 554) (95, 554) (95, 554) (97, 554) (98, 554) (99, 554) (100, 554) (101, 554) (102, 554) (103, 554) (104, 559) (106, 564) (107, 564) (148, 564) (149, 564) (150, 564) (151, 564) (153, 572) (154, 572) (156, 576) (157, 576) (158, 576) (159, 576) (160, 576) (162, 576) (164, 576) (165, 576) (166, 576) (167, 576) (169, 576) (171, 576) (172, 576) (173, 576) (174, 576) (176, 576) (178, 576) (180, 576) (181, 576) (182, 576) (184, 576) (187, 576) (189, 576) (190, 576) (192, 576) (195, 576) (196, 576) (197, 576) (198, 576) (200, 576)
				};
				
				\addplot[
					color=orange,
					mark=none
				]
				coordinates {
					(0, 37) (1, 37) (2, 112) (2, 152) (3, 152) (4, 152) (4, 152) (5, 152) (6, 152) (6, 152) (7, 152) (8, 152) (9, 152) (9, 152) (10, 152) (11, 152) (12, 152) (12, 152) (13, 152) (14, 152) (15, 152) (16, 152) (16, 152) (17, 152) (18, 152) (19, 152) (20, 152) (21, 152) (22, 152) (22, 152) (23, 152) (24, 152) (24, 152) (25, 152) (26, 152) (27, 152) (28, 152) (29, 152) (30, 152) (31, 152) (31, 152) (32, 152) (34, 152) (35, 152) (36, 152) (36, 152) (37, 152) (38, 152) (39, 152) (40, 152) (41, 152) (41, 152) (83, 152) (83, 152) (84, 152) (86, 152) (86, 152) (87, 152) (88, 152) (89, 152) (90, 152) (92, 152) (93, 152) (94, 152) (95, 152) (96, 152) (97, 152) (98, 152) (99, 152) (100, 152) (101, 152) (102, 152) (104, 152) (105, 152) (106, 152) (107, 152) (108, 152) (109, 152) (110, 152) (111, 152) (112, 152) (113, 152) (115, 152) (116, 152) (117, 152) (119, 152) (120, 152) (121, 152) (122, 152) (123, 152) (124, 152) (125, 152) (126, 152) (127, 152) (129, 152) (130, 152) (131, 152) (132, 152) (133, 152) (134, 152) (135, 152) (137, 152) (138, 152) (139, 152) (140, 152) (142, 152) (184, 152) (186, 152) (188, 152) (189, 152) (190, 152) (192, 152) (194, 152) (196, 152) (198, 152) (200, 152)
				};
				
				\legend{Total\(^{\mathrm{a}}\), Total\(^{\mathrm{b}}\), Total\(^{\mathrm{c}}\)}
			\end{axis}
		\end{tikzpicture}
	}
	\caption{RealWorld Code Coverage Trend}\label{figure:evaluation_rq1_realworld_coverage}
\end{figure}
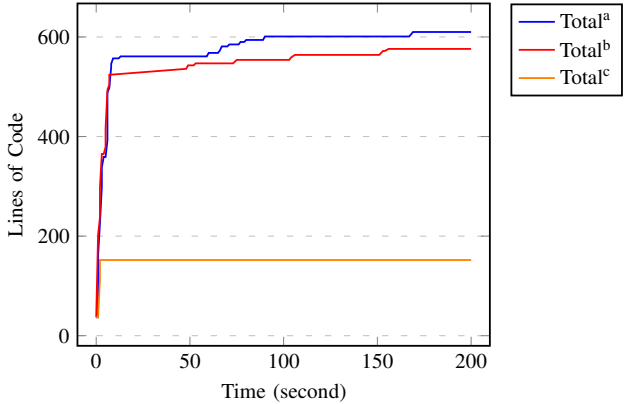

Fig.~\ref{figure:evaluation_rq1_realworld_coverage} shows that the code coverage of ``HsuanFuzz based on black box coverage level guidance'' has only been executed for about 26 rounds, and the code coverage rate increases significantly faster than ``HsuanFuzz without black box coverage level guidance''. As to RESTler, after we adjusted its ``fuzzing\_mode'' setting, whether it was bfs, bfs-cheap or random-walk, it was difficult to increase its coverage quickly.

\begin{figure}[htbp]
	\centerline{
		\begin{tikzpicture}[scale=0.8]
			\begin{axis}[
					xlabel={Time (second)}, ylabel={Lines of Code},
					xmin=-10, xmax=210,
					legend style={
						at={(1.05, 1)},
						anchor=north west,
					},
					ymajorgrids=true,
					grid style=loosely dashed,
					line width=0.3mm
				]
				\addplot[
					color=blue,
					mark=none,
					dashed
				]
				coordinates {
					(0, 21) (1, 21) (1, 48) (2, 48) (3, 48) (3, 48) (4, 48) (5, 48) (6, 48) (6, 126) (7, 136) (8, 156) (8, 156) (9, 163) (10, 163) (11, 163) (12, 163) (12, 163) (13, 163) (14, 163) (15, 163) (15, 163) (16, 163) (57, 163) (58, 163) (59, 163) (60, 163) (61, 163) (62, 163) (63, 163) (63, 163) (64, 163) (65, 163) (66, 163) (67, 163) (68, 163) (69, 163) (70, 163) (71, 163) (72, 163) (72, 163) (74, 163) (75, 163) (76, 163) (77, 163) (78, 163) (78, 163) (79, 163) (80, 163) (81, 163) (82, 163) (84, 163) (85, 163) (86, 163) (87, 163) (87, 163) (89, 163) (90, 163) (91, 163) (93, 163) (94, 163) (95, 163) (95, 163) (97, 163) (98, 163) (100, 163) (100, 163) (102, 163) (102, 163) (104, 163) (105, 163) (107, 163) (108, 163) (109, 163) (110, 163) (111, 163) (112, 163) (114, 163) (116, 163) (157, 163) (158, 163) (160, 163) (162, 163) (163, 163) (165, 163) (166, 163) (167, 163) (169, 172) (171, 172) (172, 172) (173, 172) (175, 172) (177, 172) (178, 172) (180, 172) (181, 172) (182, 172) (184, 172) (186, 172) (187, 172) (188, 172) (189, 172) (192, 172) (194, 172) (195, 172) (196, 172) (198, 172) (200, 172)
				};
				
				\addplot[
					color=red,
					mark=none,
					dashed
				]
				coordinates {
					(0, 21) (0, 21) (1, 48) (2, 48) (2, 48) (3, 48) (4, 48) (5, 48) (5, 48) (6, 126) (7, 136) (7, 156) (48, 163) (49, 163) (49, 163) (50, 163) (51, 163) (52, 163) (53, 163) (53, 163) (54, 163) (55, 163) (56, 163) (56, 163) (57, 163) (58, 163) (59, 163) (60, 163) (60, 163) (61, 163) (62, 163) (63, 163) (64, 163) (65, 163) (66, 163) (66, 163) (67, 163) (68, 163) (69, 163) (70, 163) (71, 163) (72, 163) (72, 163) (73, 163) (75, 163) (76, 163) (77, 163) (77, 163) (78, 163) (79, 163) (80, 163) (81, 163) (82, 163) (83, 163) (84, 163) (85, 163) (86, 163) (87, 163) (88, 163) (89, 163) (90, 163) (91, 163) (92, 163) (94, 163) (95, 163) (95, 163) (97, 163) (98, 163) (99, 163) (100, 163) (101, 163) (102, 163) (103, 163) (104, 163) (106, 163) (107, 163) (148, 163) (149, 163) (150, 163) (151, 163) (153, 163) (154, 163) (156, 163) (157, 163) (158, 163) (159, 163) (160, 163) (162, 163) (164, 163) (165, 163) (166, 163) (167, 163) (169, 163) (171, 163) (172, 163) (173, 163) (174, 163) (176, 163) (178, 163) (180, 163) (181, 163) (182, 163) (184, 163) (187, 163) (189, 163) (190, 163) (192, 163) (195, 163) (196, 163) (197, 163) (198, 163) (200, 163)
					
				};
				
				\addplot[
					color=orange,
					mark=none,
					dashed
				]
				coordinates {
					(0, 21) (1, 21) (2, 96) (2, 116) (3, 116) (4, 116) (4, 116) (5, 116) (6, 116) (6, 116) (7, 116) (8, 116) (9, 116) (9, 116) (10, 116) (11, 116) (12, 116) (12, 116) (13, 116) (14, 116) (15, 116) (16, 116) (16, 116) (17, 116) (18, 116) (19, 116) (20, 116) (21, 116) (22, 116) (22, 116) (23, 116) (24, 116) (24, 116) (25, 116) (26, 116) (27, 116) (28, 116) (29, 116) (30, 116) (31, 116) (31, 116) (32, 116) (34, 116) (35, 116) (36, 116) (36, 116) (37, 116) (38, 116) (39, 116) (40, 116) (41, 116) (41, 116) (83, 116) (83, 116) (84, 116) (86, 116) (86, 116) (87, 116) (88, 116) (89, 116) (90, 116) (92, 116) (93, 116) (94, 116) (95, 116) (96, 116) (97, 116) (98, 116) (99, 116) (100, 116) (101, 116) (102, 116) (104, 116) (105, 116) (106, 116) (107, 116) (108, 116) (109, 116) (110, 116) (111, 116) (112, 116) (113, 116) (115, 116) (116, 116) (117, 116) (119, 116) (120, 116) (121, 116) (122, 116) (123, 116) (124, 116) (125, 116) (126, 116) (127, 116) (129, 116) (130, 116) (131, 116) (132, 116) (133, 116) (134, 116) (135, 116) (137, 116) (138, 116) (139, 116) (140, 116) (142, 116) (184, 116) (186, 116) (188, 116) (189, 116) (190, 116) (192, 116) (194, 116) (196, 116) (198, 116) (200, 116)
				};
				
				\addplot[
					color=blue,
					mark=none
				]
				coordinates {
					(0, 16) (1, 16) (1, 112) (2, 178) (3, 251) (3, 290) (4, 311) (5, 311) (6, 344) (6, 362) (7, 362) (8, 384) (8, 390) (9, 394) (10, 394) (11, 394) (12, 394) (12, 394) (13, 398) (14, 398) (15, 398) (15, 398) (16, 398) (57, 398) (58, 398) (59, 398) (60, 405) (61, 405) (62, 405) (63, 405) (63, 405) (64, 405) (65, 405) (66, 410) (67, 418) (68, 418) (69, 418) (70, 418) (71, 422) (72, 422) (72, 422) (74, 422) (75, 422) (76, 422) (77, 427) (78, 427) (78, 427) (79, 427) (80, 431) (81, 431) (82, 431) (84, 431) (85, 431) (86, 431) (87, 431) (87, 431) (89, 431) (90, 438) (91, 438) (93, 438) (94, 438) (95, 438) (95, 438) (97, 438) (98, 438) (100, 438) (100, 438) (102, 438) (102, 438) (104, 438) (105, 438) (107, 438) (108, 438) (109, 438) (110, 438) (111, 438) (112, 438) (114, 438) (116, 438) (157, 438) (158, 438) (160, 438) (162, 438) (163, 438) (165, 438) (166, 438) (167, 438) (169, 438) (171, 438) (172, 438) (173, 438) (175, 438) (177, 438) (178, 438) (180, 438) (181, 438) (182, 438) (184, 438) (186, 438) (187, 438) (188, 438) (189, 438) (192, 438) (194, 438) (195, 438) (196, 438) (198, 438) (200, 438)
				};
				
				\addplot[
					color=red,
					mark=none
				]
				coordinates {
					(0, 16) (0, 16) (1, 154) (2, 190) (2, 250) (3, 317) (4, 317) (5, 331) (5, 368) (6, 368) (7, 368) (7, 368) (48, 373) (49, 380) (49, 380) (50, 380) (51, 380) (52, 380) (53, 384) (53, 384) (54, 384) (55, 384) (56, 384) (56, 384) (57, 384) (58, 384) (59, 384) (60, 384) (60, 384) (61, 384) (62, 384) (63, 384) (64, 384) (65, 384) (66, 384) (66, 384) (67, 384) (68, 384) (69, 384) (70, 384) (71, 384) (72, 384) (72, 384) (73, 384) (75, 391) (76, 391) (77, 391) (77, 391) (78, 391) (79, 391) (80, 391) (81, 391) (82, 391) (83, 391) (84, 391) (85, 391) (86, 391) (87, 391) (88, 391) (89, 391) (90, 391) (91, 391) (92, 391) (94, 391) (95, 391) (95, 391) (97, 391) (98, 391) (99, 391) (100, 391) (101, 391) (102, 391) (103, 391) (104, 396) (106, 401) (107, 401) (148, 401) (149, 401) (150, 401) (151, 401) (153, 409) (154, 409) (156, 413) (157, 413) (158, 413) (159, 413) (160, 413) (162, 413) (164, 413) (165, 413) (166, 413) (167, 413) (169, 413) (171, 413) (172, 413) (173, 413) (174, 413) (176, 413) (178, 413) (180, 413) (181, 413) (182, 413) (184, 413) (187, 413) (189, 413) (190, 413) (192, 413) (195, 413) (196, 413) (197, 413) (198, 413) (200, 413)
				};
				
				\addplot[
					color=orange,
					mark=none
				]
				coordinates {
					(0, 16) (1, 16) (2, 16) (2, 36) (3, 36) (4, 36) (4, 36) (5, 36) (6, 36) (6, 36) (7, 36) (8, 36) (9, 36) (9, 36) (10, 36) (11, 36) (12, 36) (12, 36) (13, 36) (14, 36) (15, 36) (16, 36) (16, 36) (17, 36) (18, 36) (19, 36) (20, 36) (21, 36) (22, 36) (22, 36) (23, 36) (24, 36) (24, 36) (25, 36) (26, 36) (27, 36) (28, 36) (29, 36) (30, 36) (31, 36) (31, 36) (32, 36) (34, 36) (35, 36) (36, 36) (36, 36) (37, 36) (38, 36) (39, 36) (40, 36) (41, 36) (41, 36) (83, 36) (83, 36) (84, 36) (86, 36) (86, 36) (87, 36) (88, 36) (89, 36) (90, 36) (92, 36) (93, 36) (94, 36) (95, 36) (96, 36) (97, 36) (98, 36) (99, 36) (100, 36) (101, 36) (102, 36) (104, 36) (105, 36) (106, 36) (107, 36) (108, 36) (109, 36) (110, 36) (111, 36) (112, 36) (113, 36) (115, 36) (116, 36) (117, 36) (119, 36) (120, 36) (121, 36) (122, 36) (123, 36) (124, 36) (125, 36) (126, 36) (127, 36) (129, 36) (130, 36) (131, 36) (132, 36) (133, 36) (134, 36) (135, 36) (137, 36) (138, 36) (139, 36) (140, 36) (142, 36) (184, 36) (186, 36) (188, 36) (189, 36) (190, 36) (192, 36) (194, 36) (196, 36) (198, 36) (200, 36)
				};
				
				\legend{User\(^{\mathrm{a}}\), User\(^{\mathrm{b}}\), User\(^{\mathrm{c}}\), Article\(^{\mathrm{a}}\), Article\(^{\mathrm{b}}\), Article\(^{\mathrm{c}}\), }
			\end{axis}
		\end{tikzpicture}
	}
	\caption{RealWorld Individual Code Coverage}\label{figure:evaluation_rq1_between_coverage}
\end{figure}
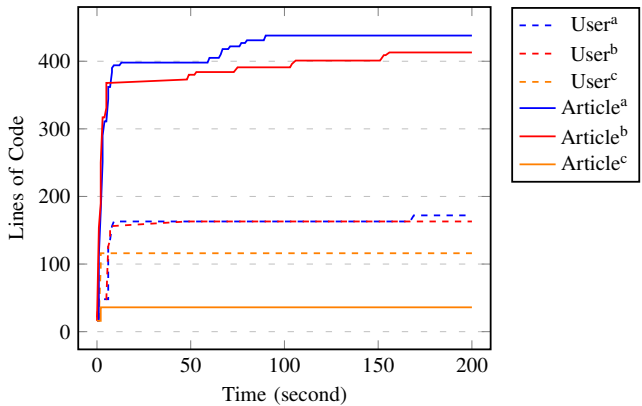

Compared with ``HsuanFuzz without black box coverage level guidelines'', ``User'' in Fig.~\ref{figure:evaluation_rq1_between_coverage} can still increase code coverage after a period of execution, while ``Article'' can increase code coverage at a faster speed. It proves that black box coverage level can guide the fuzzer effectively, converge the direction of mutation, and mutate values of parameters more quickly and properly to increase code coverage.

\subsection*{RQ 2: Service Comparisons}

We chose two open-source projects of e-commerce platform, Spree Commerce~\cite{spree} and Magento Community~\cite{magento}, as our testee and defined status code 500 as an error to evaluate the ability to find errors. The former is written in Ruby, the latter in PHP, and both are great projects on GitHub with over 9,700 stars and a combined total of over 2,300 contributors. We built Spree on version 4.2.1 and Magento on version 2.4.2 respectively in the experimental environment and entered two YAML files (path dependencies and access information) for a 24-hour test.

\subsubsection{Spree Commerce}

Spree has 31 paths and a total of 35 request methods. Table~\ref{table:evaluation_spree_bug_list} lists these three fuzzers with errors found in 24 hours.

\begin{table}[htbp]
	\caption{Errors of Spree Commerce in 24 hours}\label{table:evaluation_spree_bug_list}
	\begin{center}
		\begin{tabular}{ccccc}
			\toprule
			\textbf{Method} & \multicolumn{1}{c}{\textbf{Path}} & \textbf{H\(^{\mathrm{a}}\)} & \textbf{H\(^{\mathrm{b}}\)} & \textbf{R\(^{\mathrm{c}}\)} \\
			\midrule
			\midrule
			GET             & /order\_status/\{number\}         & \checkmark{}                &                             &                             \\
			\midrule
			GET             & /products/\{id\}                  & \checkmark{}                &                             &                             \\
			\midrule
			POST            & /account                          & \checkmark{}                & \checkmark{}                & \checkmark{}                \\
			\midrule
			POST            & /account/addresses                & \checkmark{}                & \checkmark{}                & \checkmark{}                \\
			\midrule
			POST            & /cart/add\_item                   & \checkmark{}                & \checkmark{}                & \checkmark{}                \\
			\midrule
			PATCH           & /account                          & \checkmark{}                & \checkmark{}                & \checkmark{}                \\
			\midrule
			PATCH           & /account/addresses/\{id\}         & \checkmark{}                & \checkmark{}                &                             \\
			\midrule
			PATCH           & /cart/set\_quantity               & \checkmark{}                & \checkmark{}                & \checkmark{}                \\
			\midrule
			PATCH           & /checkout                         & \checkmark{}                & \checkmark{}                & \checkmark{}                \\
			\midrule
			DELETE          & \(^{\mathrm{d}}\)                 & \checkmark{}                & \checkmark{}                &                             \\
			\midrule
			DELETE          & \(^{\mathrm{e}}\)                 & \checkmark{}                & \checkmark{}                &                             \\
			\bottomrule \\
			\multicolumn{5}{l}{\(^{\mathrm{d}}\) /cart/remove\_coupon\_code/\{coupon\_code\}}\\
			\multicolumn{5}{l}{\(^{\mathrm{e}}\) /cart/remove\_line\_item/\{line\_item\_id\}}\\
		\end{tabular}
	\end{center}
\end{table}

HsuanFuzz does not include repeated errors in its design and finds a total of 11 errors with different paths and parameters. RESTler found 55 errors. After removing duplicate errors based on the path, method, and parameters, there were a total of 6 errors.

Table~\ref{table:evaluation_spree_bug_list} shows that with a small number of paths and request methods and a long execution time, both HsuanFuzz can find a similar number of errors, but not the same number of errors, presumably due to the inability to converge the direction of mutation. RESTler could not find the error of ``/account/addresses/\{id\}'' path. We concluded that HsuanFuzz was able to find this error because it had issued the path dependency problem and adopted a good mutation strategy.

We also executed HsuanFuzz to find some additional errors. A total of 14 errors were reported for two versions of Spree~\cite{spree_issue_1}~\cite{spree_issue_2}, which were confirmed to be bugs by the project author. In addition, an error in the YAML file of OpenAPI specification was reported~\cite{spree_pr}.

\subsubsection{Magento Community}

Magento is a very large-scale service. There are 312 paths and 395 request methods in Magento Admin REST endpoints. A 24-hour test was also performed to list the errors found in these three fuzzers.

Table~\ref{table:evaluation_magento_bug_list} reveals that HsuanFuzz can find more errors than RESTler at the same time. However, RESTler cannot find an error with the request method being GET.\@ We judged that it is because the id parameter of paths will not be mutated. The number of errors found by HsuanFuzz with or without black box coverage level guidance is similar because Magento errors are mostly caused by data type changes or lack of fields, which can be triggered after fuzzing has been conducted for a certain period of time. It will be explained and compared in more detail later.

\begin{table*}[htbp]
	\caption{Errors of Magento Community in 24 hours}\label{table:evaluation_magento_bug_list}
	\begin{center}
	\begin{adjustbox}{max width=\textwidth}
		\begin{tabular}[t]{ccccc}
			\toprule
			\textbf{Method} & \multicolumn{1}{c}{\textbf{Path}}   & \textbf{H\(^{\mathrm{a}}\)} & \textbf{H\(^{\mathrm{b}}\)} & \textbf{R\(^{\mathrm{c}}\)} \\
			\midrule
			\midrule
			GET             & /\ldots \(^{\mathrm{d}}\)           & \checkmark{}                & \checkmark{}                &                             \\
			\midrule
			POST            & /\(^{\mathrm{s}}\)                  & \checkmark{}                & \checkmark{}                & \checkmark{}                \\
			\midrule
			POST            & \(^{\mathrm{e}}\)/\(^{\mathrm{w}}\) &                             &                             & \checkmark{}                \\
			\midrule
			POST            & \(^{\mathrm{f}}\)/\(^{\mathrm{t}}\) & \checkmark{}                & \checkmark{}                &                             \\
			\midrule
			POST            & \(^{\mathrm{f}}\)/\(^{\mathrm{j}}\) & \checkmark{}                & \checkmark{}                &                             \\
			\midrule
			POST            & \(^{\mathrm{f}}\)/items             & \checkmark{}                & \checkmark{}                &                             \\
			\midrule
			POST            & \(^{\mathrm{f}}\)/\(^{\mathrm{k}}\) & \checkmark{}                & \checkmark{}                &                             \\
			\midrule
			POST            & \(^{\mathrm{f}}\)/\(^{\mathrm{l}}\) & \checkmark{}                & \checkmark{}                &                             \\
			\midrule
			POST            & \(^{\mathrm{f}}\)/\(^{\mathrm{m}}\) & \checkmark{}                & \checkmark{}                &                             \\
			\midrule
			POST            & \(^{\mathrm{f}}\)/\(^{\mathrm{n}}\) & \checkmark{}                & \checkmark{}                &                             \\
			\midrule
			POST            & \(^{\mathrm{g}}\)/\(^{\mathrm{j}}\) & \checkmark{}                & \checkmark{}                &                             \\
			\midrule
			POST            & \(^{\mathrm{h}}\)/items             & \checkmark{}                & \checkmark{}                &                             \\
			\midrule
			POST            & /categories                         & \checkmark{}                & \checkmark{}                &                             \\
			\midrule
			POST            & /coupons                            & \checkmark{}                & \checkmark{}                &                             \\
			\bottomrule \\
			\multicolumn{5}{l}{\(^{\mathrm{d}}\) A total of 35 paths.}\\
			\multicolumn{5}{l}{\(^{\mathrm{e}}\) /bundle-products}\\
			\multicolumn{5}{l}{\(^{\mathrm{f}}\) /carts/mine}\\
			\multicolumn{5}{l}{\(^{\mathrm{g}}\) /carts/\{cartId\}}\\
			\multicolumn{5}{l}{\(^{\mathrm{h}}\) /carts/\{quoteId\}}
		\end{tabular}
		\begin{tabular}[t]{ccccc}
			\toprule
			\textbf{Method} & \multicolumn{1}{c}{\textbf{Path}}   & \textbf{H\(^{\mathrm{a}}\)} & \textbf{H\(^{\mathrm{b}}\)} & \textbf{R\(^{\mathrm{c}}\)} \\
			\midrule
			\midrule
			POST            & /guest-\(^{\mathrm{s}}\)            & \checkmark{}                & \checkmark{}                &                             \\
			\midrule
			POST            & \(^{\mathrm{i}}\)/\(^{\mathrm{t}}\) & \checkmark{}                & \checkmark{}                &                             \\
			\midrule
			POST            & \(^{\mathrm{i}}\)/\(^{\mathrm{j}}\) & \checkmark{}                & \checkmark{}                &                             \\
			\midrule
			POST            & \(^{\mathrm{i}}\)/items             & \checkmark{}                & \checkmark{}                &                             \\
			\midrule
			POST            & \(^{\mathrm{i}}\)/\(^{\mathrm{k}}\) & \checkmark{}                & \checkmark{}                &                             \\
			\midrule
			POST            & \(^{\mathrm{i}}\)/\(^{\mathrm{l}}\) & \checkmark{}                & \checkmark{}                &                             \\
			\midrule
			POST            & \(^{\mathrm{i}}\)/\(^{\mathrm{m}}\) & \checkmark{}                & \checkmark{}                &                             \\
			\midrule
			POST            & \(^{\mathrm{i}}\)/\(^{\mathrm{n}}\) & \checkmark{}                & \checkmark{}                &                             \\
			\midrule
			POST            & /inventory/\(^{\mathrm{o}}\)        & \checkmark{}                & \checkmark{}                &                             \\
			\midrule
			POST            & /invoices/                          & \checkmark{}                & \checkmark{}                &                             \\
			\midrule
			POST            & /orders                             & \checkmark{}                & \checkmark{}                &                             \\
			\midrule
			POST            & /products                           & \checkmark{}                & \checkmark{}                &                             \\
			\midrule
			POST            & /salesRules                         & \checkmark{}                & \checkmark{}                &                             \\
			\midrule
			POST            & /taxRates                           & \checkmark{}                &                             &                             \\
			\bottomrule \\
			\multicolumn{5}{l}{\(^{\mathrm{i}}\) /guest-carts/\{cartId\}}\\
			\multicolumn{5}{l}{\(^{\mathrm{j}}\) estimate-shipping-methods}\\
			\multicolumn{5}{l}{\(^{\mathrm{k}}\) payment-information}\\
			\multicolumn{5}{l}{\(^{\mathrm{l}}\) set-payment-information}\\
			\multicolumn{5}{l}{\(^{\mathrm{m}}\) shipping-information}\\
			\multicolumn{5}{l}{\(^{\mathrm{n}}\) totals-information}
		\end{tabular}
		\begin{tabular}[t]{ccccc}
			\toprule
			\textbf{Method} & \multicolumn{1}{c}{\textbf{Path}}               & \textbf{H\(^{\mathrm{a}}\)} & \textbf{H\(^{\mathrm{b}}\)} & \textbf{R\(^{\mathrm{c}}\)} \\
			\midrule
			\midrule
			POST            & /\(^{\mathrm{p}}\)-validation/\(^{\mathrm{p}}\) & \checkmark{}                & \checkmark{}                &                             \\
			\midrule
			PUT             & \(^{\mathrm{u}}\)/\(^{\mathrm{q}}\)             & \checkmark{}                & \checkmark{}                & \checkmark{}                \\
			\midrule
			PUT             & \(^{\mathrm{v}}\)/\(^{\mathrm{q}}\)             & \checkmark{}                & \checkmark{}                & \checkmark{}                \\
			\midrule
			PUT             & \(^{\mathrm{e}}\)/options/\{optionId\}          &                             &                             & \checkmark{}                \\
			\midrule
			PUT             & \(^{\mathrm{e}}\)/\(^{\mathrm{w}}\)             &                             &                             & \checkmark{}                \\
			\midrule
			PUT             & /carts/mine                                     &                             &                             & \checkmark{}                \\
			\midrule
			PUT             & \(^{\mathrm{f}}\)/items/\{itemId\}              & \checkmark{}                & \checkmark{}                &                             \\
			\midrule
			PUT             & \(^{\mathrm{g}}\)/items/\{itemId\}              & \checkmark{}                & \checkmark{}                &                             \\
			\midrule
			PUT             & /categories/\{id\}                              & \checkmark{}                & \checkmark{}                &                             \\
			\midrule
			PUT             & \(^{\mathrm{r}}\)/variation                     & \checkmark{}                & \checkmark{}                &                             \\
			\midrule
			PUT             & \(^{\mathrm{i}}\)/items/\{itemId\}              & \checkmark{}                & \checkmark{}                &                             \\
			\midrule
			PUT             & /orders/create                                  & \checkmark{}                & \checkmark{}                &                             \\
			\midrule
			PUT             & /products/\{sku\}                               & \checkmark{}                & \checkmark{}                &                             \\
			\bottomrule \\
			\multicolumn{5}{l}{\(^{\mathrm{o}}\) source-selection-algorithm-result}\\
			\multicolumn{5}{l}{\(^{\mathrm{p}}\) vertex-address}\\
			\multicolumn{5}{l}{\(^{\mathrm{q}}\) \{amazonOrderReferenceId\}}\\
			\multicolumn{5}{l}{\(^{\mathrm{r}}\) /configurable-products}\\
			\multicolumn{5}{l}{\(^{\mathrm{s}}\) address/cleanse}\\
			\multicolumn{5}{l}{\(^{\mathrm{t}}\) billing-address}\\
			\multicolumn{5}{l}{\(^{\mathrm{u}}\) /amazon-billing-address}\\
			\multicolumn{5}{l}{\(^{\mathrm{v}}\) /amazon-shipping-address}\\
			\multicolumn{5}{l}{\(^{\mathrm{w}}\) \{sku\}/links/\{id\}}
		\end{tabular}
		\end{adjustbox}
	\end{center}
\end{table*}

We also reported the errors we found to the project author. A total of 75 errors were reported in two versions~\cite{magento_issue_1}~\cite{magento_issue_2}, all of which were confirmed as bugs.

\subsubsection{Analysis}

After 24 hours of testing, HsuanFuzz and RESTler were able to find 11 and 6 errors out of 14 known errors for Spree, and 71 and 3 errors out of 75 known errors for Magento, proving that HsuanFuzz is superior to RESTler in terms of error finding ability.

\begin{figure}[htbp]
    \centerline{
        \begin{tikzpicture}[scale=0.8]
            \begin{axis}[
                    xlabel={Time (minute)}, 
                    ylabel={Errors},
                    legend style={
						at={(1.05, 1)},
						anchor=north west,
					},
					ymajorgrids=true,
					grid style=loosely dashed,
					line width=0.3mm
                ]
                \addplot[
                    color=blue,
                    mark=none
                ]
                coordinates {
                    (0, 0) (2, 5) (3, 7) (6, 8) (333, 9) (381, 10) (836, 11) (1440, 11)
                };

                \addplot[
                    color=red,
                    mark=none
                ]
                coordinates {
                    (0, 0) (2, 5) (3, 7)  (54, 8) (158, 9) (1440, 9)
                };

                \addplot[
                    color=orange,
                    mark=none
                ]
                coordinates {
                    (0,0)(5, 1)(24, 2)(38, 2)(59, 3)(67, 4)(68, 4)(77, 5)(1440, 5)
                };

				\legend{HsuanFuzz\(^{\mathrm{a}}\), HsuanFuzz\(^{\mathrm{b}}\) , RESTler}
            \end{axis}
        \end{tikzpicture}
    }
    \caption{Spree Commerce Errors Trend}\label{figure:evaluation_rq2_between_bugs_1}
\end{figure}
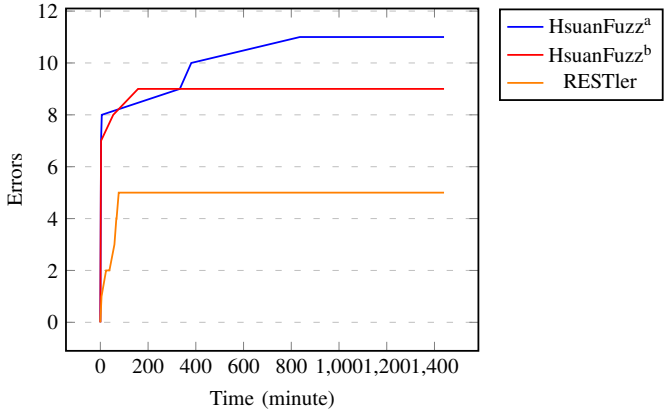

\begin{figure}[htbp]
    \centerline{
        \begin{tikzpicture}[scale=0.8]
            \begin{axis}[
                    xlabel={Time (minute)}, 
                    ylabel={Errors},
                    legend style={
						at={(1.05, 1)},
						anchor=north west,
					},
					ymajorgrids=true,
					grid style=loosely dashed,
					line width=0.3mm
                ]
                \addplot[
                    color=blue,
                    mark=none
                ]
                coordinates {
                    (0, 1) (0, 2) (0, 3) (0, 4) (0, 5) (0, 6) (0, 7) (0, 8) (0, 9) (0, 10) (0, 11) (1, 12) (1, 13) (1, 14) (1, 15) (1, 16) (1, 17) (1, 18) (1, 19) (1, 20) (1, 21) (1, 22) (1, 23) (1, 24) (1, 25) (1, 26) (1, 27) (1, 28) (1, 29) (1, 30) (1, 31) (1, 32) (1, 33) (2, 34) (2, 35) (3, 36) (3, 37) (3, 38) (3, 39) (4, 40) (5, 41) (5, 42) (5, 43) (5, 44) (6, 45) (6, 46) (8, 47) (8, 48) (8, 49) (10, 50) (10, 51) (11, 52) (12, 53) (12, 54) (13, 55) (15, 55)
                };

                \addplot[
                    color=red,
                    mark=none
                ]
                coordinates {
                    (0, 1) (0, 2) (1, 3) (1, 4) (1, 5) (1, 6) (1, 7) (1, 8) (1, 9) (1, 10) (1, 11) (1, 12) (1, 13) (1, 14) (1, 15) (1, 16) (1, 17) (1, 18) (1, 19) (1, 20) (1, 21) (1, 22) (1, 23) (1, 24) (1, 25) (1, 26) (1, 27) (1, 28) (1, 29) (1, 30) (1, 31) (1, 32) (2, 33) (2, 34) (2, 35) (3, 36) (3, 37) (4, 38) (5, 39) (5, 40) (6, 41) (7, 42) (8, 43) (8, 44) (8, 45) (8, 46) (10, 47) (10, 48) (11, 49) (12, 50) (13, 51) (13, 52) (13, 53) (13, 54) (15, 54)
                };

                \addplot[
                    color=orange,
                    mark=none
                ]
                coordinates {
                    (0,0) (15,0)
                };

				\legend{HsuanFuzz\(^{\mathrm{a}}\), HsuanFuzz\(^{\mathrm{b}}\) , RESTler}
            \end{axis}
        \end{tikzpicture}
    }
    \caption{Magento Community Errors Trend}\label{figure:evaluation_rq2_between_bugs_2}
\end{figure}
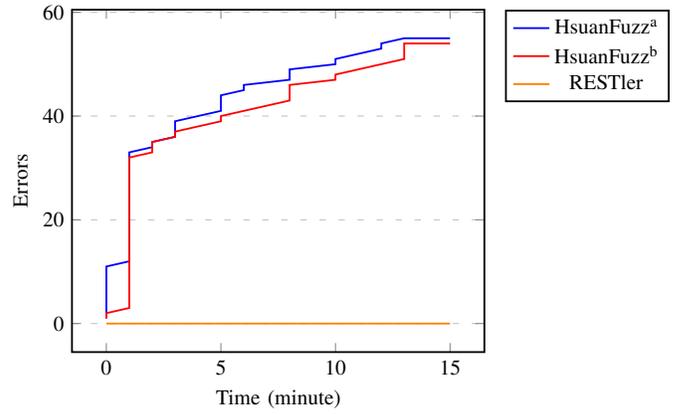

If the comparison is made from the perspective of whether ``black box coverage level guidelines'', both HsuanFuzz can find a similar number of errors in a long enough time. Therefore, we use Fig.~\ref{figure:evaluation_rq2_between_bugs_1} and Fig.~\ref{figure:evaluation_rq2_between_bugs_2} to analyze the cutting time period.

In the early stage of fuzzing, HsuanFuzz will continue to add seeds to the corpus, because there are constantly ways to improve TCL, but after a period of time, it will stop adding seeds because it cannot improve TCL.\@ However, we found that HsuanFuzz can still find new interesting seeds 2.5 hours after starting to test Spree.

Also in Fig.~\ref{figure:evaluation_rq2_between_bugs_1}, we found that it can continue to find new bugs after a long period of operation. It means that our black box coverage level guidelines are effective.

Both types of HsuanFuzz ended up with similar error results in Magento. We found that we were unable to find new errors after about two hours, and most of the error types were data type changes or missing fields, which is exactly the strategy we have defined for the REST API based on our experience. In order to give RESTler more time to find the error, we also test the same as Spree for 24 hours.

Fig.~\ref{figure:evaluation_rq2_between_bugs_2} shows the number of Magento errors found in the first 15 minutes. About 3 minutes earlier, both could find a similar number of errors, and then ``HsuanFuzz based on black box coverage level guidance'' clearly outperformed ``HsuanFuzz without black box coverage level guidance''. This shows that the feedback and guidance we added and the idea of reusing the seeds of interest are effective and that the black box fuzz testing with guidance can find errors more quickly.

\begin{figure}[htbp]
	\centerline{
		\begin{tikzpicture}[scale=0.8]
			\begin{axis}[
					ymax=90,
					ylabel= Errors,
					enlarge x limits=0.5,
					ybar,
					xtick=data,
					symbolic x coords={Spree Commerce, Magento Community},
					ymajorgrids=true,
					bar width=0.8cm,
					nodes near coords,
					legend pos=north west,
				]
				\addplot coordinates { (Spree Commerce, 11) (Magento Community, 71)};
				\addplot coordinates { (Spree Commerce, 9) (Magento Community, 70)};
				\addplot coordinates { (Spree Commerce, 6) (Magento Community, 3)};
				\legend{HsuanFuzz\(^{\mathrm{a}}\), HsuanFuzz\(^{\mathrm{b}}\) , RESTler}
			\end{axis}
		\end{tikzpicture}
	}
	\caption{Errors in 24 hours}\label{figure:evaluation_rq2_bugs_compare}
\end{figure}
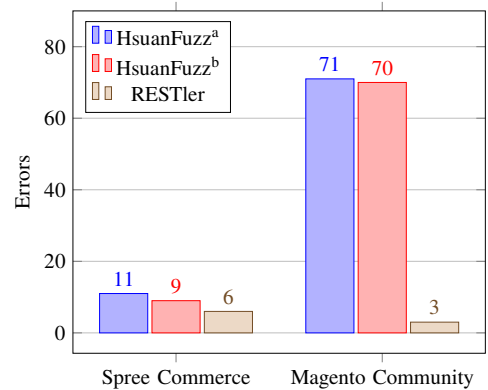

According to Fig.~\ref{figure:evaluation_rq2_bugs_compare}, HsuanFuzz has a good error finding ability, while Fig.~\ref{figure:evaluation_rq2_between_bugs_1} and Fig.~\ref{figure:evaluation_rq2_between_bugs_2} show that the coverage level guide is effective in finding new interesting seeds after a certain period of time. Finally, Table~\ref{table:evaluation_rq2_error_type} shows the type and number of errors we reported.

\begin{table}[htbp]
	\caption{Type and Number of Errors}\label{table:evaluation_rq2_error_type}
	\begin{center}
	\begin{adjustbox}{max width=\columnwidth}
		\begin{tabular}{ccc}
			\toprule
			\textbf{Target} & \multicolumn{1}{c}{\textbf{Type}}                  & \textbf{Number} \\
			\midrule
			\midrule
			Spree           & No Method Error                                    & 7               \\
			\midrule
			Spree           & SQL Exception                                      & 2               \\
			\midrule
			Spree           & Invalid URI                                        & 2               \\
			\midrule
			Spree           & Unknown Attribute                                  & 1               \\
			\midrule
			Spree           & Page Incalculable                                  & 1               \\
			\midrule
			Spree           & Type Error                                         & 1               \\
			\midrule
			Spree           & Column not found                                   & 27              \\
			\midrule
			Magento         & Property \(A\) does not have accessor method \(B\) & 26              \\
			\midrule
			Magento         & Missing required argument    \(A\)                 & 7               \\
			\midrule
			Magento         & Failed to parse time string                        & 3               \\
			\midrule
			Magento         & Amazon Pay could not process your request          & 2               \\
			\midrule
			Magento         & Request name \(A\) doesn't exist                   & 2               \\
			\midrule
			Magento         & \(A\) must be of type string                       & 2               \\
			\midrule
			Magento         & Undefined offset    \(N\)                          & 2               \\
			\midrule
			Magento         & Invalid argument supplied for foreach ()           & 1               \\
			\midrule
			Magento         & Expects parameter \(N\) to be array                & 1               \\
			\midrule
			Magento         & Class \(A\) does not exist                         & 1               \\
			\midrule
			Magento         & Invalid scope type \(A\)                           & 1               \\
			\bottomrule
		\end{tabular}
		\end{adjustbox}
	\end{center}
\end{table}

\subsection*{RQ 3: Other Advantages}

The grey box fuzzer requires program instrumentation and has the drawback of programming language restrictions. For black box testing, it does not need to understand the internal behavior and the programming language used. Therefore, black box fuzzing can be performed on a large scale, quickly and effectively without having to set up a Web API server and without having background knowledge of the target service's programming language.

APIs.guru~\cite{apis-guru} is committed to becoming a Web API Wikipedia, and only public, persistent and useful Web APIs will be included. It provides many different types of services, most of which can be accessed directly, and some can be browsed only with subscription or authorization. With the specifications of Swagger and OpenAPI versions, it can provide a perfect entry point for black box testing.

We selected 936 YAML files with OpenAPI, a total of 768 services, and tested them without entering path dependencies and access information. Finally, 465 responses with a status code greater than ``500'' were found in 64 services, including 351 with the status code of ``500 Internal Server Error'' in 47 services.

We succeeded in finding the error of the remote target service without knowing the API programming language. By collecting relevant information in the response, it can be found that the target service uses programming languages such as Ruby, PHP, C\#, and Java, which also proves that black box testing has the advantage of not being restricted by languages.

\begin{table*}[htbp]
	\caption{Test Results of Remote APIs}\label{table:evaluation_rq3_bug_list}
	\begin{center}
		\begin{tabular}[t]{lc}
			\toprule
			\multicolumn{1}{c}{\textbf{Test Target}} & \textbf{500s} \\
			\midrule
			\midrule
			rest.zuora.com                           & 3             \\
			\midrule
			api.logoraisr.com/rest-v1                & 3             \\
			\midrule
			apirest.isendpro.com/cgi-bin             & 1             \\
			\midrule
			rest.ably.io                             & 3             \\
			\midrule
			ibl.api.bbci.co.uk/ibl/v1                & 1             \\
			\midrule
			apps.gov.bc.ca/pub/bcgnws                & 1             \\
			\midrule
			api.beezup.com                           & 2             \\
			\midrule
			bikewise.org/api                         & 3             \\
			\midrule
			www.bungie.net/Platform                  & 110           \\
			\midrule
			rest-api.d7networks.com/secure           & 2             \\
			\midrule
			dev.to/api                               & 5             \\
			\midrule
			api.presalytics.io/doc-converter         & 1             \\
			\midrule
			api.open511.gov.bc.ca                    & 1             \\
			\midrule
			exude-api.herokuapp.com                  & 1             \\
			\midrule
			api.figshare.com/v2                      & 10            \\
			\midrule
			apps.gov.bc.ca/pub/geomark               & 4             \\
			\bottomrule
		\end{tabular}
		\begin{tabular}[t]{lc}
			\toprule
			\multicolumn{1}{c}{\textbf{Test Target}}  & \textbf{500s} \\
			\midrule
			\midrule
			apps.nrs.gov.bc.ca/gwells/api/v1          & 1             \\
			\midrule
			oralquestionsandmotions-api.parliament.uk & 1             \\
			\midrule
			api.interzoid.com                         & 1             \\
			\midrule
			go.netlicensing.io/core/v2/rest           & 6             \\
			\midrule
			marketcheck-prod.apigee.net/v2            & 51            \\
			\midrule
			nsidc.org/api/dataset/2                   & 3             \\
			\midrule
			v2.namsor.com/NamSorAPIv2                 & 2             \\
			\midrule
			ntp1node.nebl.io                          & 17            \\
			\midrule
			www.neowsapp.com                          & 7             \\
			\midrule
			api.oceandrivers.com                      & 5             \\
			\midrule
			osdb.openlinksw.com/osdb                  & 1             \\
			\midrule
			demo.orthanc-server.com                   & 10            \\
			\midrule
			api.paylocity.com/api                     & 23            \\
			\midrule
			peertube2.cpy.re/api/v1                   & 1             \\
			\midrule
			phantauth.net                             & 5             \\
			\midrule
			api.pocketsmith.com/v2                    & 5             \\
			\bottomrule
		\end{tabular}
		\begin{tabular}[t]{lc}
			\toprule
			\multicolumn{1}{c}{\textbf{Test Target}} & \textbf{500s} \\
			\midrule
			\midrule
			randomlovecraft.com/api                  & 2             \\
			\midrule
			connect.squareup.com                     & 5             \\
			\midrule
			api.stoplight.io/v1                      & 3             \\
			\midrule
			api.parliament.uk/search                 & 2             \\
			\midrule
			api.up.com.au/api/v1                     & 4             \\
			\midrule
			api.icons8.com                           & 1             \\
			\midrule
			api.sandbox.velopayments.com             & 4             \\
			\midrule
			vocadb.net                               & 18            \\
			\midrule
			worldtimeapi.org/api                     & 2             \\
			\midrule
			ws.api.video                             & 1             \\
			\midrule
			api.ideaconsult.net/nanoreg1             & 3             \\
			\midrule
			geodesystems.com                         & 2             \\
			\midrule
			api.cloudmersive.com                     & 12            \\
			\midrule
			smart-me.com                             & 1             \\
			\midrule
			api.spoonacular.com                      & 1             \\
			\bottomrule
		\end{tabular}
	\end{center}
\end{table*}

\subsection*{Threats To Validity}
   Greybox fuzz testing has been proved to be effective in discovering security bugs. However, programs either in source or binary form need to be instrumented for coverage feedback. We propose to take advantage of TCL (test coverage level) in blackbox testing to be as a replacement of conventional code coverage. It is the primary threat to our work. Since TCL is only an approximation of test coverage, it relies on the response and request through the OpenAPI specifications. Currently, there are only seven levels of TCL and it is too coarse to be as precise as the code coverage.
\section{Related Work}\label{chapter:related_work}

RESTler~\cite{RESTler} is currently a well-known REST API black box fuzzer, proposed by Atlidakis et al.\ and as Microsoft's open-source project~\cite{restler_github}. RESTler is a generation-based fuzzer that uses a predefined dictionary to replace parameter values in grammar and analyzes OpenAPI to infer dependencies between paths statically. It is to lengthen the test case as much as possible with most paths for a single round of requests. In contrast, our research uses manual input to achieve a more accurate path dependency. We also add TCL feedback to guide the fuzzer and increase the chance of triggering errors through mutation-based fuzzing and other strategies.

Pythia~\cite{Pythia} is based on grey box fuzzing, also proposed by Atlidakis et al., using the built-in functions of Ruby to get code coverage. The initial seed is the execution result of RESTler, and the mutation stage is improved by the machine learning method, and the validity of the grammar is still maintained after some noise is injected. Coverage-guided fuzzing in Python-based web server~\cite{shihtsunliu} is modified from Python-AFL, which is also one of the grey box fuzzing methods, which can obtain code coverage from Flask. The grey box fuzzing methods mentioned above have limitations, and the test targets are limited by the programming language of the fuzzer. 

EvoMaster has two methods: white box fuzzing~\cite{EvoMaster2017} and black box fuzzing~\cite{EvoMaster2021}. The former needs to read the document and write part of Java code for the test target in order to generate a test case. It requires programming ability. The latter is tested randomly,  to maximize types of HTTP status codes, and does not have an appropriate feedback mechanism.

QuickREST~\cite{QuickREST} is a property-based black box testing method that can quickly verify and test OpenAPI but does not send a request to the test target, so it may be different from the developer's implementation result. RestTestGen~\cite{RESTTESTGEN} is also a black box test method, after executing an effective test, try to trigger an error by violating restrictions and other methods. In addition, customized rules such as renaming fields or stemming are used to analyze the interdependence of paths. In our research, the target is continuously tested, and the strategy of adding variant strings increases the chance of triggering errors, and the accuracy of path dependencies is improved through manual input.
\section{Conclusions}\label{chapter:conclusion}

This research has developed a proof-of-concept tool, a REST API black box fuzzer based on coverage level guidelines. The purpose is to produce more appropriate test cases for black box fuzzing. In the case of avoiding incorrect requests, meaningfully mutations are conducted to generate new values and test every path as much as possible.

We add test coverage levels feedback to guide the fuzzer, and resolves the drawback of black box testing that cannot know the effect of the mutation. We also resolve the issues of testing complexity for REST APIs and strengthen mutation strategy to increase the chance of triggering errors. 

Our research is inspired by AFL which provides feedback based on code coverage. We uses input and output of specifications to approximate code coverage by test coverage levels. We enter the appropriate OpenAPI, path dependencies, and authorization information to automatically generate test cases and perform black box fuzzing on the test target API.\@ To our best knowledge, this is the first research to apply test coverage levels for REST API black box fuzzing.

It is important to note that this research is not a grey box test, and does not require instrumentation of program or server. One of the benefits of using black box testing is that it is not restricted by any programming language and does not need to understand the internal behavior of the program. Through the OpenAPI that clearly defines the input and output, we can test the remote target API more easily.\@ Black box fuzzing with feedback guidance currently seems to be the best way to test a large number of REST APIs.

This study proves that the coverage level guidelines are effective in the estimated form of code coverage, and also proves that both self-built and remote test targets can effectively find errors. Finally, in two large-scale open-source projects, a total of 89 errors were found and reported, and 351 errors were also found in 64 remote services provided by APIs.guru.

\subsection*{Future Work}

\subsubsection{Increasing Coverage Level} 

There are a total of 8 test coverage levels used in this research, ranging from level 0 to 7, which can be measured from multiple aspects of the API, and the relative relationship with code coverage can be obtained. If more levels are added, whether the test coverage can be more accurate and easier to find errors, is worth further research and discussion.

\subsubsection{Manual Analysis and Automatic Identification}

When recording errors of the remote test target, we found that some servers would return version information, such as ``Tomcat 8.5.60'' and ``Tomcat 9.0.39''. However, these versions have CVEs (CVE-2021--25122, CVE-2021--25122, and CVE-2021--25329).

We can use this research as the first stage to conduct large-scale testing. After finding out the possible weaknesses or vulnerabilities in the system, the second stage of manual analysis or automatic identification can be carried out to reduce the cost of manually verifying the test targets one by one.

\subsubsection{Supporting IoT and CoAP}

CoAP has a REST style and has the advantages of low power consumption. It is designed to be used on devices with relatively insufficient resources to meet the needs of the Internet of Things and the requirements for the use of networked devices.

Since it is similar to HTTP, it is worth further study by modifying the test coverage standard and using it on CoAP.\@ We expect to extend this research to CGI, IoT communication protocol to identify weaknesses and try to find out the vulnerabilities of firmware components.

\section*{Acknowledgment}

This work was supported in part by the Ministry of Science and Technology
(MOST107-2221-E-009-030-MY3 and MOST-110-2218-E-A49-011-MBK), and the Cybersecurity Center of Excellence project at Taiwan’s National Applied Research Labs.
%\balance
%\ifpdf
% \IEEEtriggeratref{15}
%\fi
\bibliographystyle{IEEEtran}
\bibliography{references}
%\balance
%\flushend
\end{document}